\documentclass[twocolumn]{aastex631}
\usepackage{amsmath}
\usepackage{xspace}
\usepackage{threeparttable}
\DeclareUnicodeCharacter{00A0}{ }

\newcommand{\nova}{V1674~Her\xspace}

\newcommand{\xmm}{{\it XMM-Newton}\xspace}

\newcommand{\swift}{{\it Swift}\xspace}

\newcommand{\nustar}{\textit{NuSTAR}\xspace}

\newcommand{\fluxcgs}{erg\,cm$^{-2}$\,s$^{-1}$\xspace}
\newcommand{\lumcgs}{erg~s$^{-1}$\xspace}
\newcommand{\lat}{{\it Fermi}-LAT\xspace}
\newcommand{\xrt}{{\it Swift}-XRT\xspace}
\begin{document}

\title{X-ray spectroscopy mass constraints on \nova: the fastest nova does not have a near-Chandrasekhar white dwarf}

\author[0009-0005-3149-4327]{Tekeba Olbemo}
\affiliation{Department of Physics, Washington University in St. Louis, St. Louis, 
MO 63130, USA}
\author[0000-0002-1853-863X]{Manel Errando}
\affiliation{Department of Physics, Washington University in St. Louis, St. Louis, 
MO 63130, USA}

\author[0000-0002-5726-5216]{Andrea Gokus}
\affiliation{Department of Physics, Washington University in St. Louis, St. Louis, 
MO 63130, USA}

\correspondingauthor{Tekeba Olbemo}
\email{n.olbemo@wustl.edu}

\begin{abstract}
\noindent \nova (Nova Her 2021) is the fastest classical nova ever recorded, with an optical decline time of $t_2 \sim 1$ day, typically interpreted as evidence for a white dwarf mass close to the Chandrasekhar limit. We present a broadband X-ray study of \nova combining contemporaneous \xmm and \nustar observations in quiescence to directly constrain the white dwarf mass and magnetic field strength. 

\noindent The hard X-ray emission is modeled using a physically motivated post-shock accretion column model that accounts for the temperature gradient in the flow and reflection from the white dwarf surface. Under the assumption that the accretion disk is truncated at the co-rotation radius, we obtain a white dwarf mass of $M = 1.09^{+0.07}_{-0.06}\,M_\odot$. An independent constraint derived from timing analysis of the X-ray power spectrum yields a consistent value of $M = 1.12 \pm 0.06\,M_\odot$. 

\noindent These values are significantly lower than those inferred from empirical decline-time relations, suggesting that such relations may overestimate white dwarf masses in extreme fast novae. From the inferred accretion rate and magnetospheric radius, we estimate a surface magnetic field strength of $B = 21.3^{+6.6}_{-5.7}\,(\mathrm{stat})^{+12.9}_{-8.1}\,(\mathrm{sys})\,\mathrm{MG}$, placing \nova at the high end of the magnetic field distribution for intermediate polars. 

\noindent Our results demonstrate that even the fastest novae do not necessarily host near-Chandrasekhar white dwarfs, highlighting the importance of direct X-ray constraints and suggesting that additional parameters beyond white dwarf mass play a key role in setting nova timescales.
\end{abstract}

\keywords{Intermediate Polars, DQ Herculis stars, Cataclysmic variable stars, X-ray binary stars, White dwarf stars, Classical novae, Novae}

\section{Introduction} \label{sec:intro}
\nova, also known as Nova Her 2021 and TCP~J18573095+1653396, is a classical nova discovered by Seiji Ueda, Kushiro, Hokkaido, Japan, on 2021 June 12.548 UT at a visual magnitude of  8.4.\footnote{\url{http://www.cbat.eps.harvard.edu/unconf/followups/J18573095+1653396.html}}.
The nova was observed at high-cadence in the hours leading to the optical peak, capturing an initial slow rise and a momentary dip at $\sim14.7$\,mag followed by a fast rise to the maximum optical brightness at visual magnitude of 6.2 \citep{2024ApJ...977...17Q}. The multiwavelength emission was monitored across the electromagnetic spectrum,  from radio to gamma rays \citep{2021ApJ...922L..10W,2021ApJ...922L..42D,2022MNRAS.517L..97L,2023MNRAS.521.5453S,2024MNRAS.528...28B}. 

\nova is also classified as an intermediate polar based on the observed periodic modulations in its optical and X-ray light curves with spin period of $\sim 8.4$\,min \citep{2021ATel14720....1M,2021ATel14776....1M,2021ApJ...922L..42D,2022ApJ...932...45O,2024MNRAS.528...28B} and orbital period of $\sim0.153$\,d \citep{2021JAVSO..49..257S,2021PZ.....41....4S,2021ATel14856....1P,2022MNRAS.517L..97L,2024BAAA...65...60L}. Intermediate polars are a sub-category of cataclysmic variables that contain magnetic white dwarfs. They have magnetic fields  strong enough to disrupt the inner regions of the accretion disk \citep{1995cvs..book.....W}. The observed spin period modulation in intermediate polars is likely a result of self-occultation by the white dwarf and/or the phase-dependent absorption in the accretion column \citep{2008MNRAS.387.1157R}. 
\nova belongs to the currently small group of magnetic novae observed with modern X-ray instruments, the other members being V4743~Sgr, V2491~Cyg and V407~Lup \citep{2022ApJ...932...45O}.

Based on the observed change between the white dwarf spin periods before and after the nova eruption, and assuming that it is due to loss of angular momentum in the outburst,  \cite{2021ApJ...922L..42D} estimated the mass of the ejecta in the 2021 \nova explosion to be in the range from $2\times10^{-5}$ to $2\times10^{- 4}$\,M\textsubscript{\(\odot\)}. \cite{2021ApJ...922L..10W} calculated an upper limit on the hydrogen ejecta mass of $1.4^{+0.8}_{-1.2}\times10^{-3}$ M\textsubscript{\(\odot\)} from near-infrared spectroscopy of the nova on day 11. Detailed photo-ionization modeling of \nova optical spectra at several epochs (days 10 to 28) by \cite{2024MNRAS.527.1405H} indicates ejecta mass in the range of $3.42\,\mbox{--}\,7.04\times10^{-5}$M\textsubscript{\(\odot\)}.  \cite{2023MNRAS.521.5453S}, on the other hand, estimated a rather low ejecta mass of $10^{-7}$ M\textsubscript{\(\odot\)} using the radio emission and the lack of intrinsic absorption observed in the X-ray spectrum of the nova. 

 \begin{figure*}[tb]
  \centering
  \includegraphics[scale=0.5]{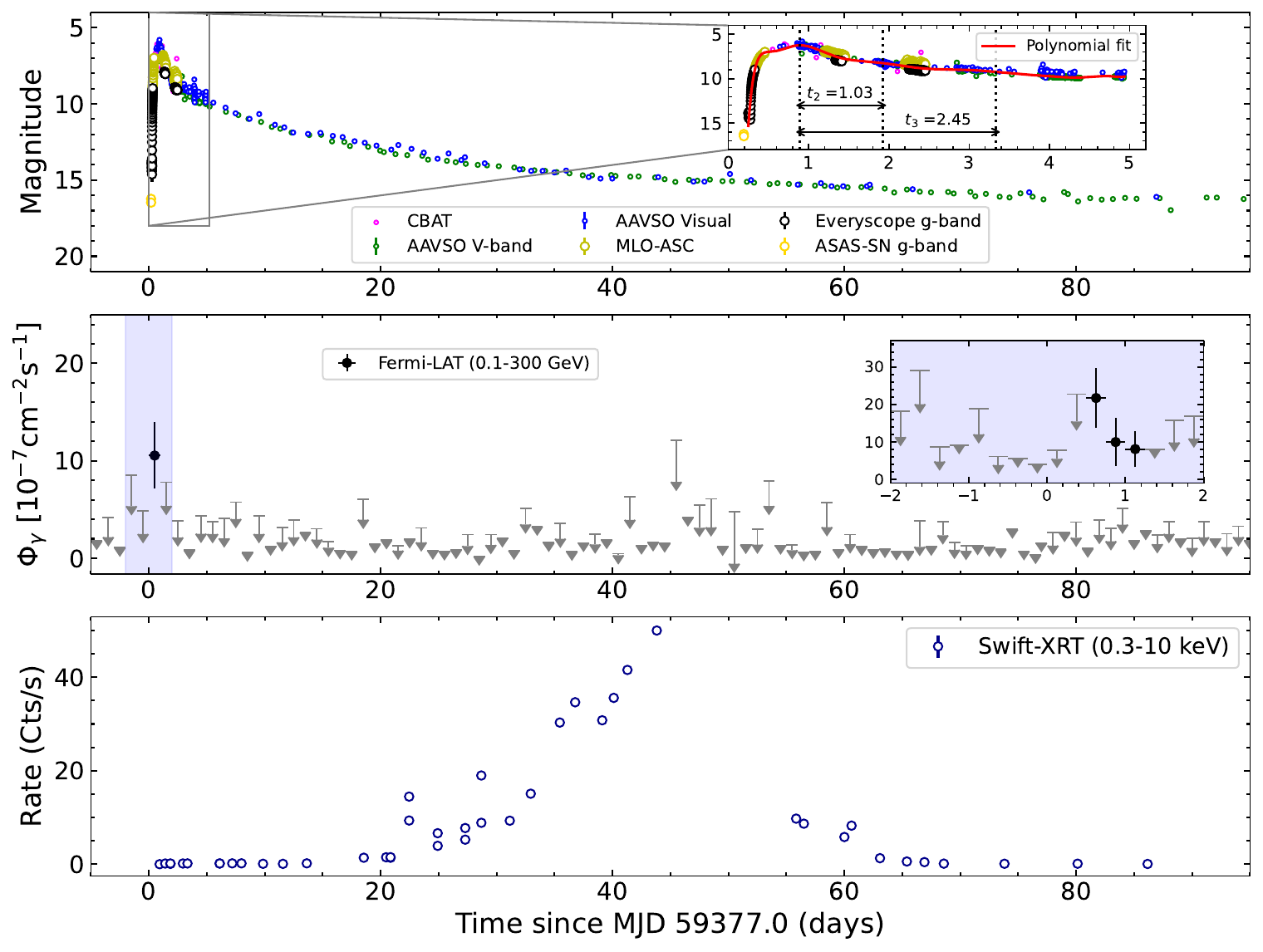}
\caption{{\it Top:} Optical ({\it top}), \lat (({\it middle})) and \xrt (({\it bottom})) light curves of \nova. Inset in the top panel show estimates for the optical decline times, $t_2$ and $t_3$ from polynomial fits to the optical light curve during the nova outburst. Inset in the middle panel shows 6\,h binned Fermi-LAT light curve around the time of the optical peak.}
 \end{figure*}
 \label{fig:lc1}

Among the unique features of \nova is its extremely rapid optical decline, with $t_2\sim 1.0$\,days \citep[][Figure~\ref{fig:lc1}]{2024ApJ...977...17Q}, defined as the time it takes for the optical brightness to fade by two magnitudes after reaching its maximum. This makes \nova the fastest-declining nova ever recorded. Such an unusually fast decline is generally interpreted as evidence for a very small ignition mass and, consequently, a high surface gravity on the white dwarf.  
Under otherwise similar accretion conditions, this strongly suggests that the white dwarf mass must be close to the Chandrasekar limit.
However, the mass of a white dwarf cannot exceed the Chandrasekhar limit of $M\sim 1.4\,M_{\odot}$, because the electron degeneracy pressure, the force that supports the star against gravitational collapse, can only sustain hydrostatic equilibrium up to this mass. 

In this work, we provide the first broadband X-ray constraint on both the white dwarf mass and magnetic field of \nova, combining physically motivated post-shock modeling with X-ray timing diagnostics.

\section{X-ray and gamma-ray observations and data reduction} \label{sec:xrt}
\subsection{\xrt}
\xrt observed \nova about 55 times between 2021-06-13 and 2022-09-23 for a total duration of $\sim$65
ks. These observations, which monitored the nova at different phases of the outburst, included exposures recorded in Windowed Timing (WT) and Photon Counting (PC) modes. Data from these observations were reduced using \texttt{xrtpipeline} version 0.13.7 on \texttt{HEASOFT} version 6.32. Only grade 0 events were considered for the light curve and spectral extraction.

Spectra for each individual observation were extracted following the approach described in \cite{2009MNRAS.397.1177E}. The approach prescribes extracting ancillary response files (ARFs) for individual snapshots and using the count weighted sum of these ARFs for the total source spectrum extracted for a given observation. To obtain a source spectrum per observation, we first divided the events (where applicable) into snapshots and considered them possibly piled up if the PC (WT) count rate in the time interval exceeded 0.6 (150) counts/s. For events with count rates below this limit, we used the circular source extraction region of radius $R\leq70\arcsec$  centered at the position of the source \nova:
$\alpha(\mathrm{J2000}) = 18\fh57\fm30\fs98,\ 
\delta(\mathrm{J2000}) = +16^\circ 53\arcmin39\arcsec6$.

About 20\% of all observations during the outburst are affected by pile up. For piled up events (PC/WT modes), an annular extraction region of outer radius 70\arcsec centered at the source’s coordinates was used. The inner radius of the annular extraction region for PC mode events was determined by fitting the XRT point spread function (PSF) model to the outer wings of the PSF of observed data. For WT mode events the smallest inner radius that ensures the observed source count rate below the above mentioned pile up limit of 150 counts/s was selected. ARFs were created using \texttt{xrtmkarf} task with PSF correction included to correct for the loss of counts due to annular source extraction region(if piled up) and bad pixels. We used an annular region with inner radius 150\arcsec and outer radius 322\arcsec centered at the nova coordinates for background extraction. For WT mode spectra, the BACKSCAL keywords in the source and background spectrum files were adjusted to reflect the ratio in radial (1D) sizes of respective extraction regions. Similarly RESPFILE keywords were set to point to the appropriate response files, ‘swxwt0s6\textunderscore20210101v017.rmf’ (WT) and ‘swxpc0s6\textunderscore20210101v016.rmf’ (PC) in our case. Lastly, spectra were  grouped with grppha command group min 1 to ensure that there are no bins the with zero counts before fitting the spectrum using Cash statistic. \texttt{xspec} version 12.13.1e  \citep{1996ASPC..101...17A} was used for spectral fitting. 

\subsection{XMM-Newton}
We extract data taken with the pn detector \citep{struder2001} of the European Photon Imaging Camera (EPIC) onboard \xmm \citep{jansen2001} between 0.3 and 10\,keV. In particular, we use the data obtained on October 14 2023 (ObsID 0931400101) and version 21.0.0 of the XMM Science Analysis software (XMMSAS).
We extract a spectrum and a light curve with 10s binning based on standard methods to process observation data files and produce a calibrated event list as well as an image.  
The EPIC-pn observation contained a total exposure time of 55\,ks and was performed in full window mode.
We extract only single and double events, and use circles for the source and background regions with a radius of 35'' and 60'', respectively. The background region is chosen with sufficient distance to the \nova in a source-free region.
We find that intense background activity coincided with the observation and we remove data in affected time ranges from our analysis. As a result, the usable exposure time is reduced to $\sim18$\,ks.

\subsection{NuSTAR}
\nustar observed \nova 6 days after the nova outburst on 23 June 2021 (obsID: 90701321002), and once again two years later during quiescence on 14 October 2023 (obsID: 30901021002) for a total exposure time of 80 ks. Data from the \nustar two telescope detector modules FPMA and FPMB were reduced using \nustar Data Analysis Software (\texttt{NuSTARDAS}) v1.5.2 and \texttt{HEASOFT} v6.7. Level 2 event files were generated using \texttt{nupipeline} task. Source photons were extracted from a circular region of radius $50\arcsec$ centered at the optical coordinates of the nova. Four circular regions of radius $50\arcsec$ from source-free regions were used to obtain background events. The \texttt{nuproducts} task was used to extract spectra and light curves. Barycentric correction was applied to the photon arrival times before extracting light curves. Spectra on the \nustar energy band ( 3-78\,keV) was extracted and grouped with \texttt{grppha} tool to contain at least 20 counts per bin for chi-squared fit statistic.

\subsection{\lat}
We used \texttt{fermipy} version 1.4.0 with ScienceTools version 2.2.0 \citep{2017ICRC...35..824W} for binned analysis of the \lat data. We used Pass 8 data and P8R3\_SOURCE\_V3 instrument response functions. We applied data selection to include only the source class photons (evclass=128, evtype=3) with good quality (DATA\_QUAL\textgreater0 \&\& LAT\_CONFIG=1). Zenith angle is restricted to a maximum of 90$^{\circ}$ to avoid contamination from the Earth's limb. 
We considered events within a region of interest (ROI) of 15$^{\circ}$ radius centered at the optical coordinate of \nova. We included photons with the reconstructed energy range of 0.1--300 GeV.  Our source model comprises all 4FGL catalog sources located inside a 25$^{\circ}$ circle from \nova, as well as the galactic (gll\_iem\_v07.fits) and the isotropic (iso\_P8R3\_SOURCE\_V3\_v1.txt) diffuse emission components. 
In order to model the long term emission of sources close to our target, we fitted the $\sim$ 1 year data between 2020-06-01 and 2021-06-01 (before the nova outburst) without including \nova in the source model. Spectral parameters of galactic and isotropic diffuse emission were free in the model. Spectral parameters of point sources are left to vary in the fit if they are within 5$^{\circ}$ from the target source and are detected with $\geq 10\sigma$ significance in 4FGL-DR3 source catalog. We also left normalizations free for variable sources in the catalog (variability index greater than 24.73) if located within  10$^{\circ}$ from the target source. The best fit spectral parameter values from this initial fit are fixed and then used in the subsequent fit including our target source, \nova, based on LAT data collected around the nova outburst (between days $\sim T_{0}-10$ and T$_{0}+100$). We extracted a light curve for this time period with bin sizes of 1 day. We also calculated light curves with finer time bins of 0.25 days (6.0 hours) for a brief time period around the optical peak.  

\section{Analysis of the nova light curve} \label{sec:xlc}
We adopt the date of nova eruption on MJD 59377.0 (2021 June 12.0) as a reference time (T$_0$). Figure~\ref{fig:lc1} shows the evolution of the X-ray light curve after the nova outburst from \nova, along with the optical and \lat gamma ray light curves.

From a polynomial fit to the optical light curve including data from ASASN-SN \citep{2014AAS...22323603S,2023arXiv230403791H}, Evryscope and MLO-ASC \citep{2014SPIE.9145E..0ZL, 2021RNAAS...5..160Q}, CBAT and AAVSO \citep{zz} observations, we obtained a decline time scale of $t_2=1.03$ days confirming the speed class of \nova as the fastest decaying nova. This is in close agreement with the range of $t_2$ values for the nova reported in the literature \citep{2021RNAAS...5..160Q, 2021PZ.....41....4S, 2024MNRAS.527.1405H, 2024ApJ...977...17Q}.  The daily binned LAT light curve of the nova indicates short-lived gamma-ray  activity with a single peak on 12 June 2021. The 6 hours binned LAT light curve around this time (the optical peak) shows $\geq 2\sigma$ LAT detections (test statistic, TS $\gtrsim 6.0$ for two degrees of freedom) between days T$_{0}+0.5$ and T$_{0}+1.25$. This gamma-ray emission and peak time window overlaps with the time period of the optical peak, possibly indicating a contribution from shocks to the optical luminosity of the nova. Previous observations of correlated optical and gamma-ray flares during nova outbursts have revealed strong evidence for shock-powered optical emission \citep{2020NatAs...4..776A}.  
The peak X-ray emission (0.3--10 keV) from V1674 Her appeared approximately  45 days after the nova outburst. 

\begin{figure*}[tb]
  \centering
  \includegraphics[scale=0.7]{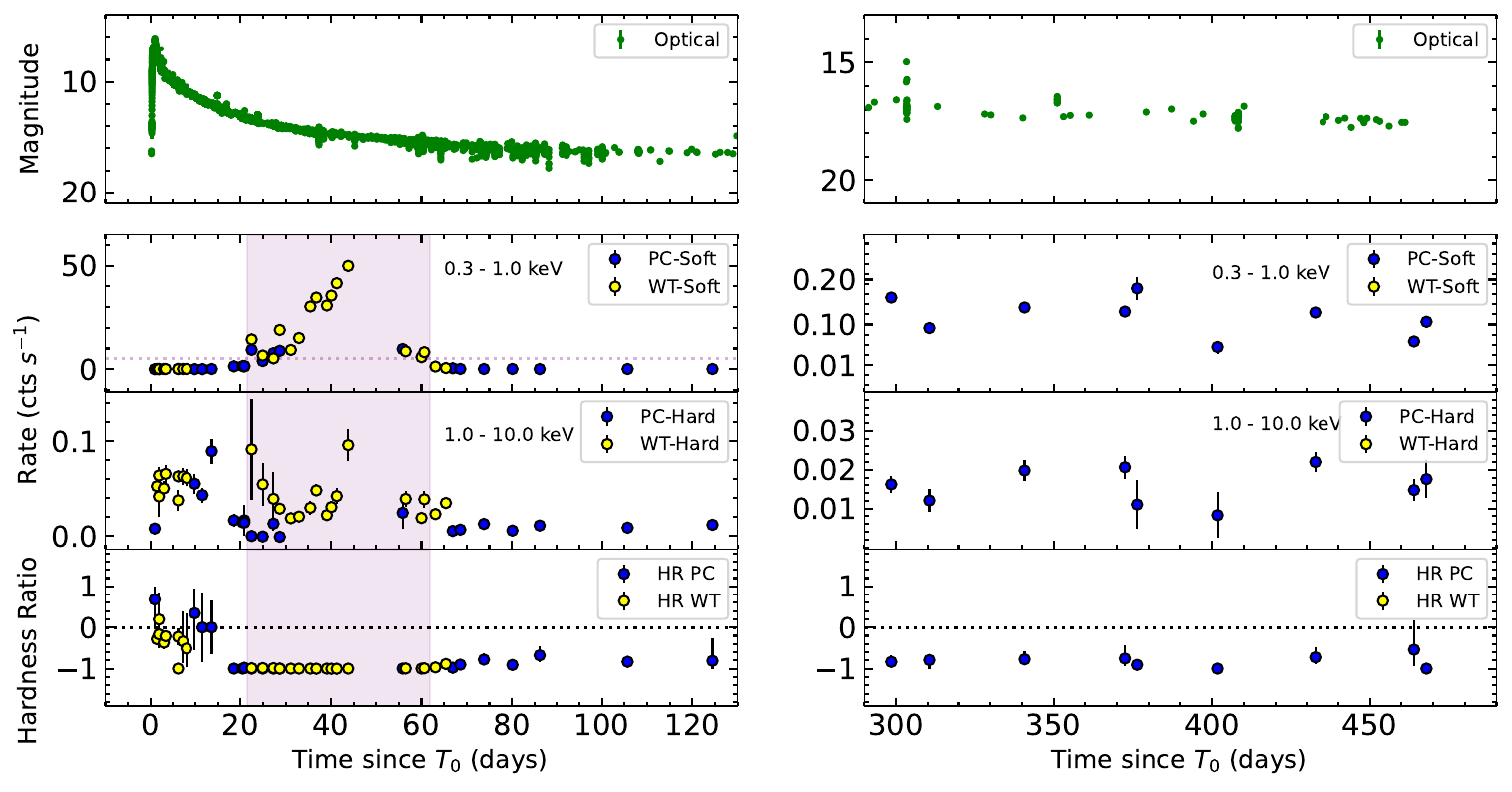}
\caption{Swift-XRT light curve evolution of V1674 Her in outburst (left) and quiescence (right). Optical light curves are included for comparison. Pink shaded region indicate the super-soft  phase. }
   \label{fig:lc2}
 \end{figure*}

Detailed \xrt X-ray light curves are depicted in Figure~\ref{fig:lc2}, showing separate light curves generated for a soft ($S$, 0.3--1.0\,keV) and a hard ($H$, 1.0--10.0\,keV) band, as well as the resulting hardness ratio. As shown in the light curves, some initial exposures around the optical peak and most observations near the X-ray peak, during the super-soft source phase, were taken in the WT readout mode to mitigate  optical loading and pile-up issues, respectively. The hardness ratio ($HR$) figure are calculated as
\begin{gather}
\label{eq:1}
HR = \frac{H-S}{H+S}\\
\sigma_{HR}=\frac{2}{(H+S)^2}\sqrt{H^2 \sigma_S^2+S^2 \sigma_H^2}
\end{gather}
 
For observations with a small number of photon counts(less than 10 photon counts), we used a similar Bayesian approach of $HR$ and uncertainty calculation described in  \cite{13cf25a983094d89970443b0ba89be68}.

As shown in Figure~\ref{fig:lc2}, the initial X-ray flux is faint and is dominated by a hard spectral component. The soft X-ray emission component takes over and dominates the observed radiative output around three weeks after the nova eruption, beginning super-soft source phase on the nova.  \citet{2020PASJ...72...82K} define this time window as the period where the X-ray emission first rises from one tenth of its peak value, through the peak flux, and then declines back to one tenth of its peak value. In \nova, we observe that the super-soft  phase lasted about 40 days, from day 21.5 to day 61.8 after $T_0$ (see the pink shaded region in Figure~\ref{fig:lc2}). 
\citet{2022ATel15312....1W} reported that the \nova outburst was over by $T_0 + 288$ days, indicating that the system had entered quiescence nine months after the outburst.
\xrt observed the nova several times in the post-outburst quiescent phase. The emission in this period was relatively weak, with an average count rate of 0.1\,cts/s. The soft band emission on average was an order of magnitude greater than the hard X-ray emission, still dominating the total radiative output. In quiescent state, the X-ray emission originates from the accretion column on the intermediate polar white dwarf rather than the nova outflow that dominates the post-outburst X-ray flux. 

\begin{figure*}[tb]
  \centering
  \includegraphics[scale=0.45]{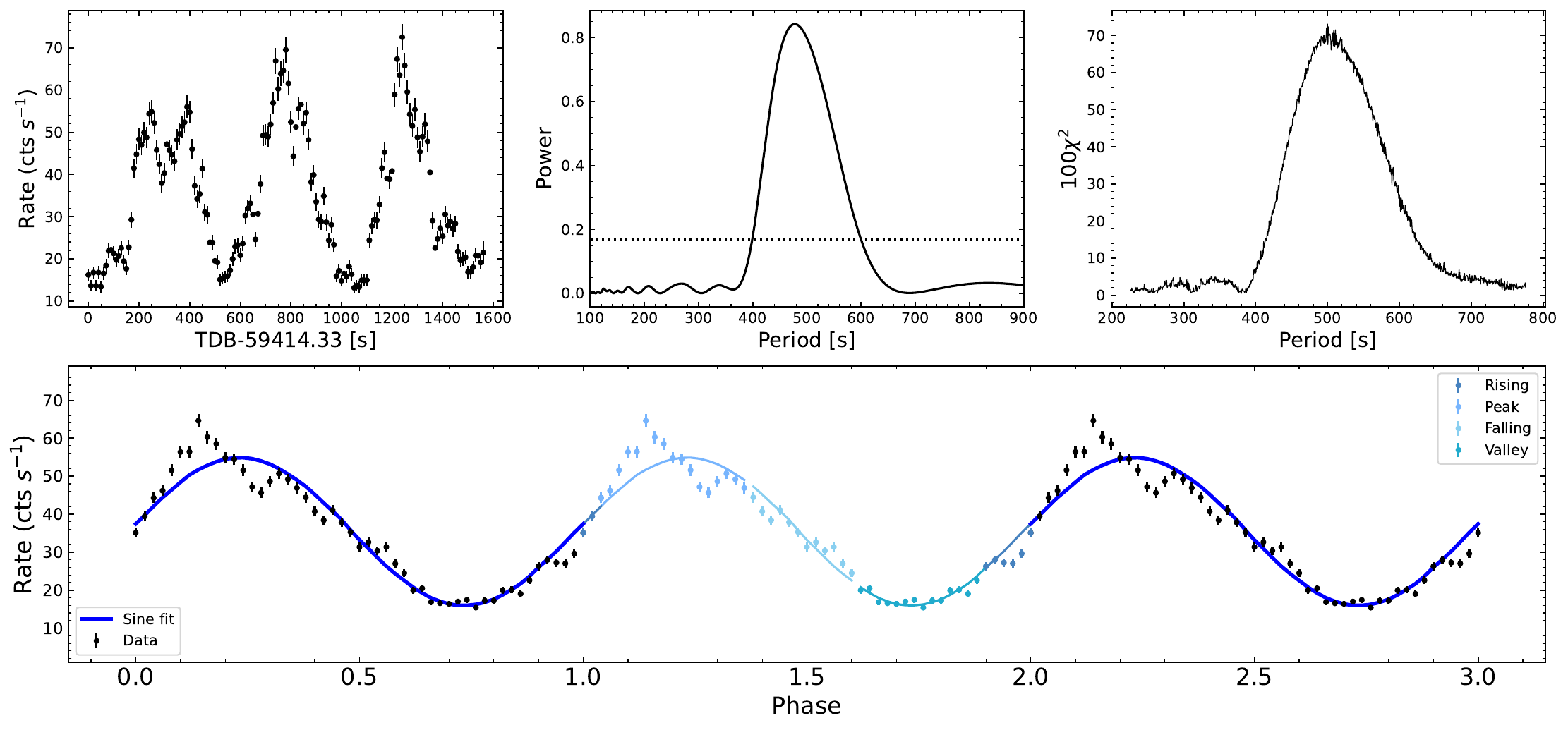}
  \includegraphics[scale=0.45]{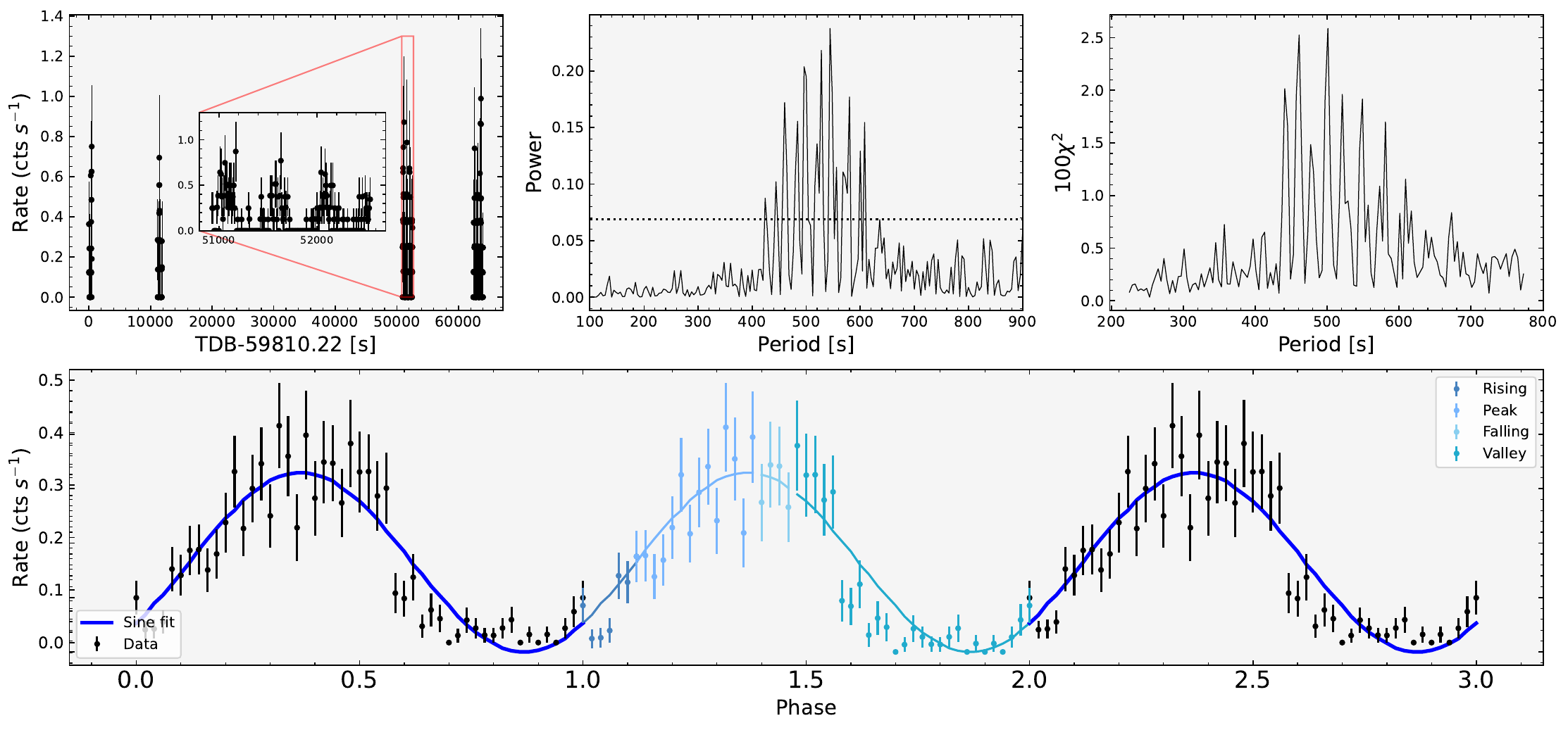}
\caption{White dwarf spin modulated X-ray emission in the outburst (white background) and quiescence (gray background). Panels 1 and 3 show the light curve (left), LS periodogram (center) and epoch folding period search results (right). Panels 2 and 4 show phase folded light curves and a simple sinusoidal fit to the data. Different colors in phase folded light curve plots indicate different white dwarf spin phases.}
   \label{fig:lc3}
 \end{figure*}

We extracted 10\,s binned light curves for $\sim 19$ individual \xrt observations with an exposure greater than 1\,ks to search for the periodic modulations reported in optical and other X-ray observations. We searched for periodicity by calculating the Lomb-Scargle periodogram  \citep{1976Ap&SS..39..447L,1982ApJ...263..835S}, as well as using the epoch folding technique \citep{1983ApJ...266..160L}. Examples of an outburst (19 Jul 2021) and quiescence (19 Aug 2021) period searches are shown in Figure \ref{fig:lc3}. We detect significant periodicity in both light curves (peaks of periodograms exceed the 0.1\% false alarm probability). The epoch folding peaks at $\sim 500$\,s are close to the white dwarf spin period values measured using other X-ray and optical observations of the nova \citep{2021ApJ...922L..42D,2021ATel14856....1P,2022ApJ...932...45O,2024MNRAS.528...28B}. 

\begin{figure*}[tb]
  \centering
  \includegraphics[scale=0.7]{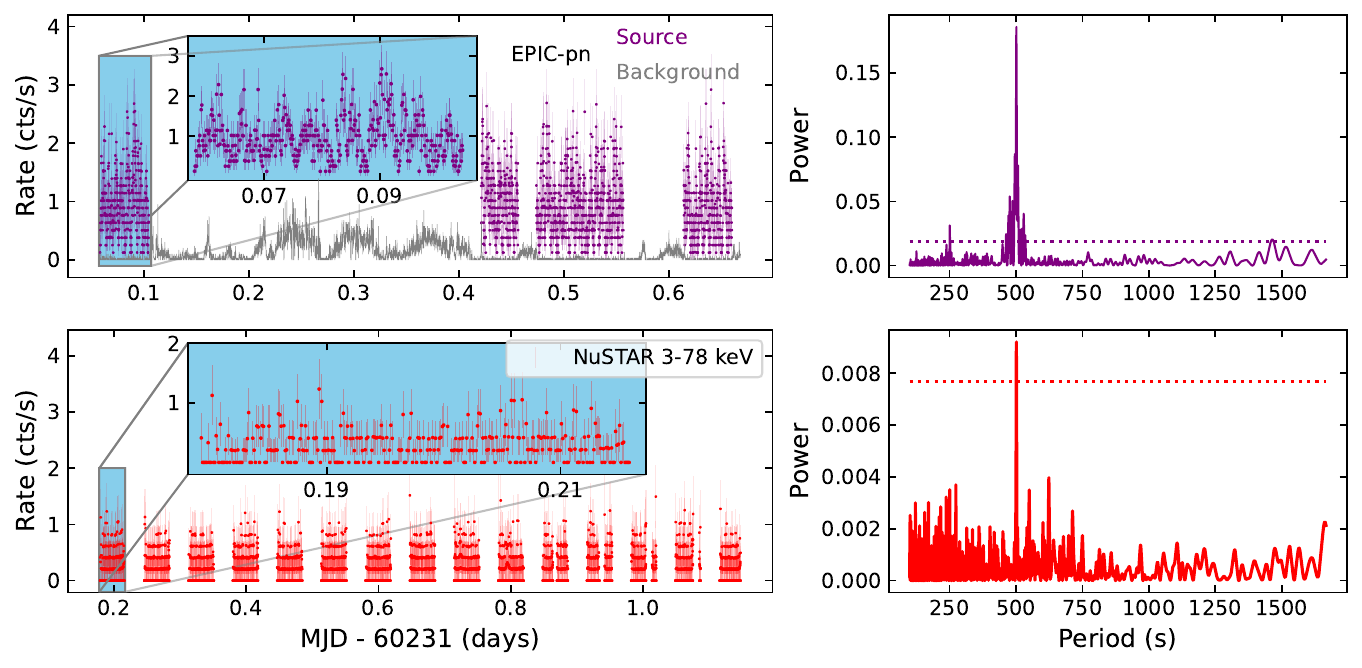}
\caption{XMM-Newton/PN (top left) and NuSTAR (bottom left) light curve of V1674 Her. Corresponding Lomb-Scargle periodograms for XMM-Newton/PN (top right) and NuSTAR (bottom right) light curves. Horizontal dotted lines indicate FAP=0.1\%.}
   \label{fig:lc_xmm_Nu}
 \end{figure*}

We also extracted 10\,s binned light curves from the \xmm and \nustar data obtained on \nova on 14 October 2023 (Figure~\ref{fig:lc_xmm_Nu}). Periodicity analysis using the Lomb-Scargle periodogram yields significant periodicity with corresponding period values of $501.73\pm 0.080$\,s and $501.4\pm 0.56$\,s, respectively, which is also in close agreement with previously reported values of the white dwarf spin period.
 
\section{Results from X-ray spectroscopy} \label{spc}
\begin{figure*}[tb]
  \centering
  \includegraphics[scale=0.7]{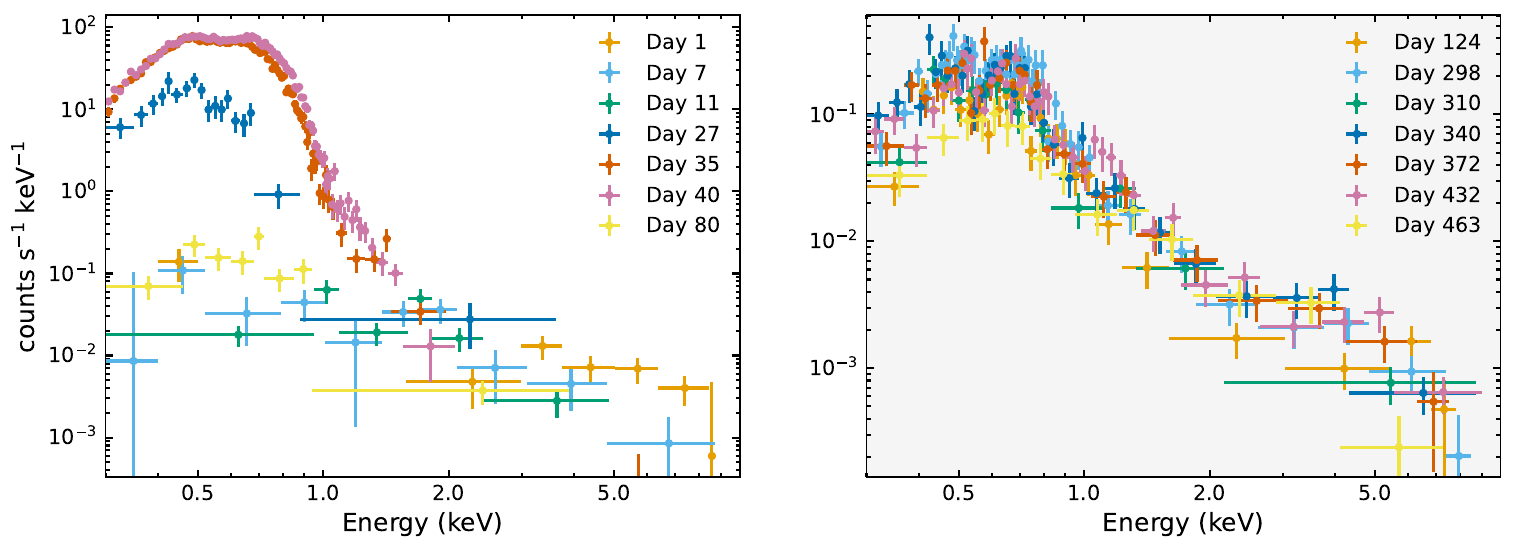}
\caption{Swift-XRT spectra of V1674 Her in outburst (left) and quiescence (right).}
   \label{fig:XRT_spec}
 \end{figure*}

The 0.3--10\,keV \xrt spectra of \nova observed at different epochs in Figure~\ref{fig:XRT_spec} show faint hard-spectrum X-ray emission in the first two weeks after the nova outburst followed by a super-soft spectral component that dominates the X-ray flux for the next 60 days or so. The hard X-ray emission observed in the early phase of the nova outburst is attributed to optically thin emission from the shock heated gas \citep{2001A&A...373..542O}. 

\begin{figure*}[tb]
  \centering
  \includegraphics[scale=0.6]{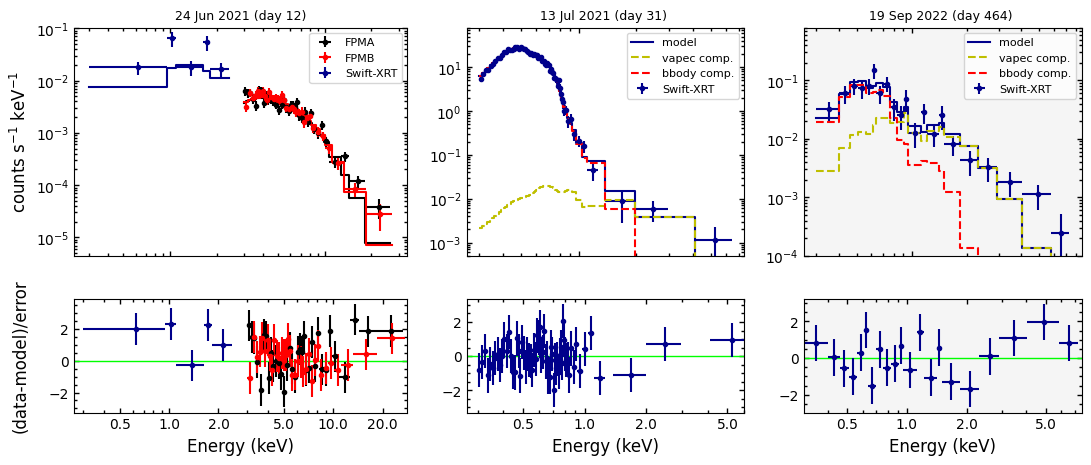}
\caption{The early shock (left), the super-soft source (middle) and the quiescent (right) phase X-ray spectra of V1674 Her. }
   \label{fig:NuXRT_spec}
 \end{figure*}

We fit the quasi-simultaneous \xrt and \nustar spectra of the nova obtained 12 days after the outburst (Figure~\ref{fig:NuXRT_spec}, left panel) with an absorbed  \texttt{apec} model with solar abundance gas \citep{2001ApJ...556L..91S}, indicating a plasma temperature  $kT=7.2^{+0.5}_{-0.4}$\,keV ($\chi^2/\text{d.o.f} = 301.2/69$). 
Using the same model with gas abundances obtained from spectroscopy of the brightest gamma-ray nova in the GeV band, V906~Car \citep{2020MNRAS.497.2569S}, results in a lower plasma temperature of $kT=4.1\pm 0.2$\,keV with improved fit statistics of $\chi^2/\text{d.o.f} = 87.1/69$ . These best fit values are in agreement with the preferred fit results reported in \cite{2023MNRAS.521.5453S}, indicating ejecta with non-solar composition.  

\begin{figure*}[tb]
  \centering
  \includegraphics[scale=0.7]{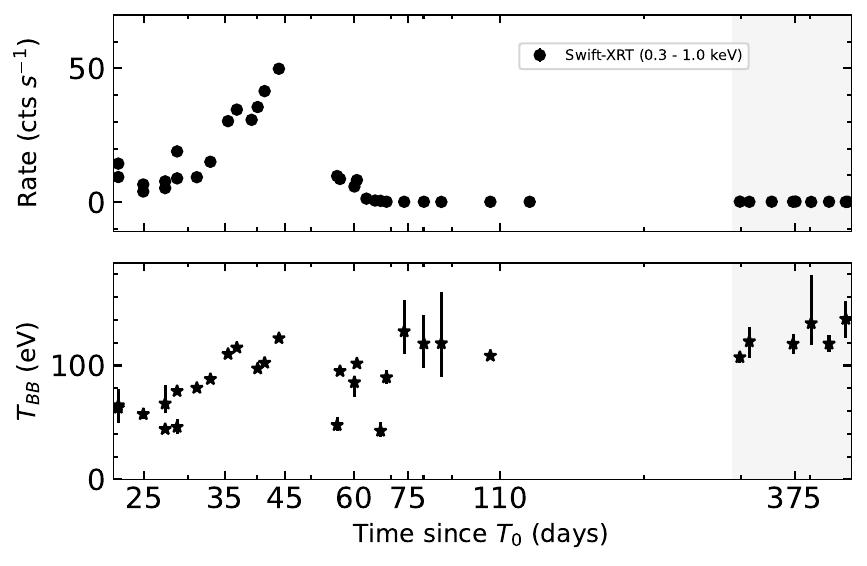}
\caption{Evolution of the X-ray flux (Top panel) and black body temperature (Bottom panel) }
   \label{fig:bb}
 \end{figure*}

The residual nuclear burning of the accreted hydrogen rich envelope on the surface of the white dwarf contributes to the delayed super-soft  phase emission observed in the nova explosions. In the early phase of the outburst, the emission originating from the stellar surface of the white dwarf is highly attenuated by the strong absorbing column of the optically thick ejecta. As the ejecta expands and thins out, the opacity decreases, revealing the soft emission from nuclear burning on the white dwarf envelope. 
This emission can be described with a black body model. A fit to a \xrt spectrum of \nova obtained 21 days after the outburst (dominated by super-soft emission) is shown in Figure~\ref{fig:NuXRT_spec} (middle panel). An absorbed \texttt{bbody+apec} model including absorption edges corresponding to ionized H-like and He-like oxygen ({O\,VIII}, {O\,VII}) and nitrogen ({N\,VII}, {N\,VI}) provides a good fit to the observed spectrum. The addition of these absorption edges significantly improves the fit and reduces the  residuals. As the post-outburst system continues to evolve, the X-ray spectra indicate black body temperatures gradually increasing from $\sim$50\,eV to peak values exceeding 100\,eV (Figure~\ref{fig:bb}). The temperature of the radiating black body stays at these relatively high values even after the total X-ray flux significantly drops.
This could be due to inner hot nuclear burning sites or heated polar cap areas from intermediate polar accretion becoming accessible as the photosphere recedes further towards the white dwarf envelope.

The right panel of Figure~\ref{fig:NuXRT_spec} shows the same absorbed \texttt{bbody+apec} model fit to the XRT spectra of \nova in a quiescent state, 464 days after the outburst.  
Intermediate polars, unlike polars, have magnetic fields of an intermediate strength ($\sim 0.1–10$\,MG) that are not strong enough to completely disrupt the accretion disk. The accretion disk in intermediate polars, instead, is truncated at the radius where the accreted matter begins to free-fall onto the white dwarf surface along the magnetic field lines. This results in a supersonic flow that forms a standing shock above the white dwarf surface. In the post-shock region, the gas eventually cools by emitting primarily hard X-ray radiation. The soft X-ray emission observed in several intermediate polars is attributed to the heating of the polar caps at the base of the accretion flow \citep{2007ApJ...663.1277E}. This component is described by a black body model. The black body model component fits to the spectra of \nova in quiescent state show temperatures around 100\,eV (Figure~\ref{fig:bb}).  
These temperatures exceed those inferred during the nova outburst but are consistent with the blackbody temperatures observed in intermediate polars, which span a range from $\sim$30 to $\sim$120\,eV \citep{2008A&A...489.1243A}.

\begin{table*}[t!]
\centering
\caption{Phase resolved spectra fit results}
\label{table:1}
\begin{tabular}{ll ccccc}
\hline
Epoch & Phase & $n_{\text{H}}^a$ & $kT^b$ & $norm^{b,*}$ & $\tau^c$ & $\tau^d$ \\
 & & [$10^{22}$ cm$^{-2}$] & [keV] & & & \\
\hline
19 Jul 2021 & Rise & $0.42^{+0.05}_{-0.07}$ & $70.0^{+15.7}_{-20.0}$ & $3.3^{+48.8}_{-1.7}$ & $3.5^{+0.2}_{-0.5}$ & $7.6^{+1.7}_{-1.2}$ \\
            & Peak   & $0.35^{+0.02}_{-0.02}$ & $104.9^{+3.2}_{-3.3}$  & $0.3^{+0.1}_{-0.1}$  & $2.5^{+0.1}_{-0.1}$ & $8.8^{+0.8}_{-0.8}$ \\
            & Fall & $0.31^{+0.03}_{-0.02}$ & $110.2^{+2.5}_{-5.5}$  & $0.1^{+0.1}_{-0.0}$  & $2.8^{+0.2}_{-0.2}$ & 10.0$^{\dagger}$ \\
            & Valley  & $0.38^{+0.03}_{-0.05}$ & $99.8^{+5.3}_{-3.7}$   & $0.2^{+0.1}_{-0.1}$  & $2.0^{+0.3}_{-0.2}$ & 10.0$^{\dagger}$ \\
\hline
19 Aug 2022 & Rise & $0.29^{+0.16 \dagger}$ & $107.8^{+12.6}_{-26.6}$ & $2.3^{+35.2}_{-1.1}$ & $1.4^{+1.4}_{-1.4}$ & $1.3^{+2.7}_{-1.0}$ \\
            & Peak   & $0.29^{+0.03 \dagger}$ & $118.6^{+8.4}_{-8.7}$   & $3.1^{+2.1}_{-0.9}$  & $1.0^{+0.4}_{-0.4}$ & 0.0$^{\dagger}$ \\
            & Fall & $0.29^{+0.22 \dagger}$ & $124.1^{+22.8}_{-22.3}$ & $2.7^{+6.7}_{-1.5}$  & $2.2^{+1.5}_{-2.0}$ & 0.0$^{\dagger}$ \\
            & Valley  & $0.36^{+0.12}_{-0.06}$ & $113.1^{+21.4}_{-16.6}$ & 1.4$^{\dagger}$      & $3.0^{+1.0}_{-0.9}$ & 0.0$^{\dagger}$ \\
\hline
\multicolumn{7}{l}{$^a$ Galactic absorption; $^b$ Blackbody component; $^*$ $\times 10^8$ for 19 Jul 2021, $\times 10^4$ for 19 Aug 2022;} \\
\multicolumn{7}{l}{MaxTau for absorption edge at $^c$ 0.74 keV; $^d$ 0.87 keV; $^{\dagger}$ Parameter pegged at hard limit} \\
\end{tabular}
\end{table*}

\subsection{Phase Resolved Spectra}

\begin{figure*}[tb]
  \centering
  \includegraphics[scale=0.65]{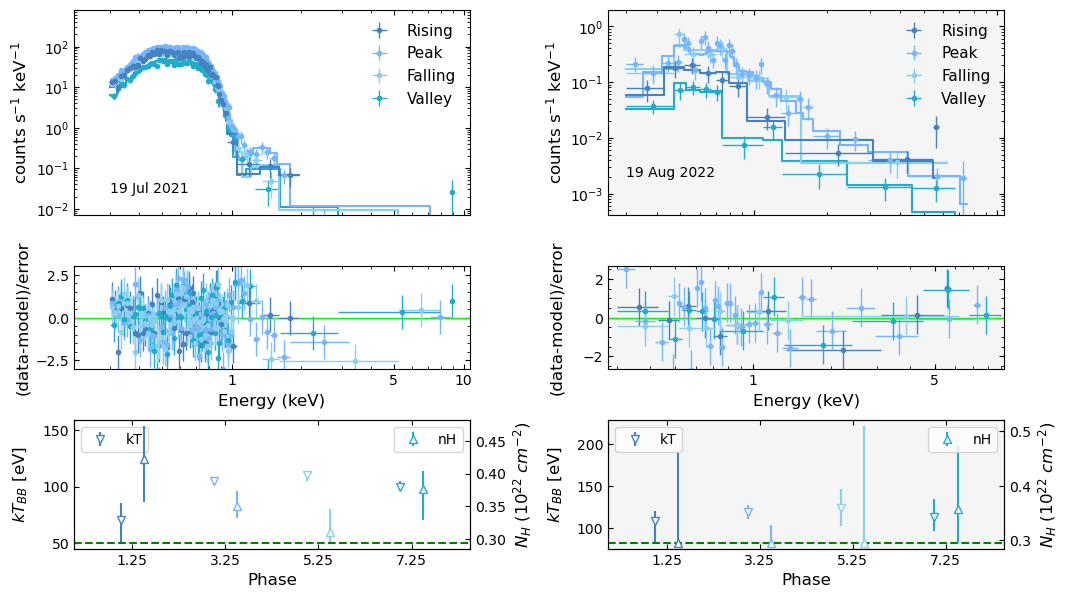}
\caption{Phase resolved spectra in the outburst (left) and quiescence (right): {\it Top:} Spectra {\it Middle:} Residuals and {\it Bottom:} Blackbody temperature and Absorption column density. Horizontal dashed green lines indicate the galactic line-of-sight hydrogen column density of $N_{\mathrm{HI}}=2.94\times10^{21}\,cm^{-2}$\citep{2016A&A...594A.116H}.}
   \label{fig:phRspec}
 \end{figure*}

A defining property of intermediate polars is that their X-ray emission is modulated at the spin period of the white dwarf. 
In \nova, this periodicity is evident in \xrt, \xmm and \nustar light curves as shown in Figures~\ref{fig:lc3} and \ref{fig:lc_xmm_Nu}. We extracted phase resolved \xrt spectra for two representative observations during the outburst and quiescence states 
to search for changes in the X-ray spectrum correlated with the spin phase of the white dwarf. We extracted spectra in four distinct phases:rise, peak, decay, and valley. 
Spectra were fit with the absorbed \texttt{bbody+apec} model mentioned above. The fit results depicted in Figure~\ref{fig:phRspec} and Table~\ref{table:1} indicate that there is no significant variation in spectral shape between different phases. 
These results favor a scenario in which a combination of inhomogeneous emission and self-occultation by the white dwarf body is a more likely cause of the spin modulated X-ray emission in \nova \citep[see also][]{2021ApJ...922L..42D}.

\begin{table*}[t!]
\centering
\caption{Model components and significance}
\label{table:2}
\begin{tabular}{l *{5}{L}}
\hline
 Component & $\chi^2_\nu/dof$ &$\Delta\chi^2$ &F&p-value&\\
 \hline
\texttt{edge}&1.43/781&299.91&142.74&$1.11\times10^{-16}$&\\
\texttt{pcfabs}&1.07/781&14.27&6.79&$1.19\times10^{-3}$&\\
\texttt{reflect}&1.06/781&10.22&4.86&$7.97\times10^{-3}$&\\
\texttt{bbody}&6.01/781&3877.5&1845.53&$1.11\times10^{-16}$&\\
\texttt{gauss}&1.07/781&19.47&9.27&$1.05\times10^{-4}$&\\
\hline
\end{tabular}
\end{table*}
\subsection{Joint XMM-Newton and NuSTAR spectral analysis}\label{sec:spcbremss}
\xmm and \nustar obtained contemporaneous observations of the quiescent X-ray emission from \nova on 14 October 2023.
The hard X-ray emission from intermediate polars can be roughly described by bremsstrahlung emission from shock heated gas falling onto the surface of the white dwarf \citep[e.g.][]{1984MNRAS.211..883K,1994PASP..106..209P,1995A&A...297L..37H}.

    \begin{figure*}[tb]
  \centering
  \includegraphics[scale=0.7]{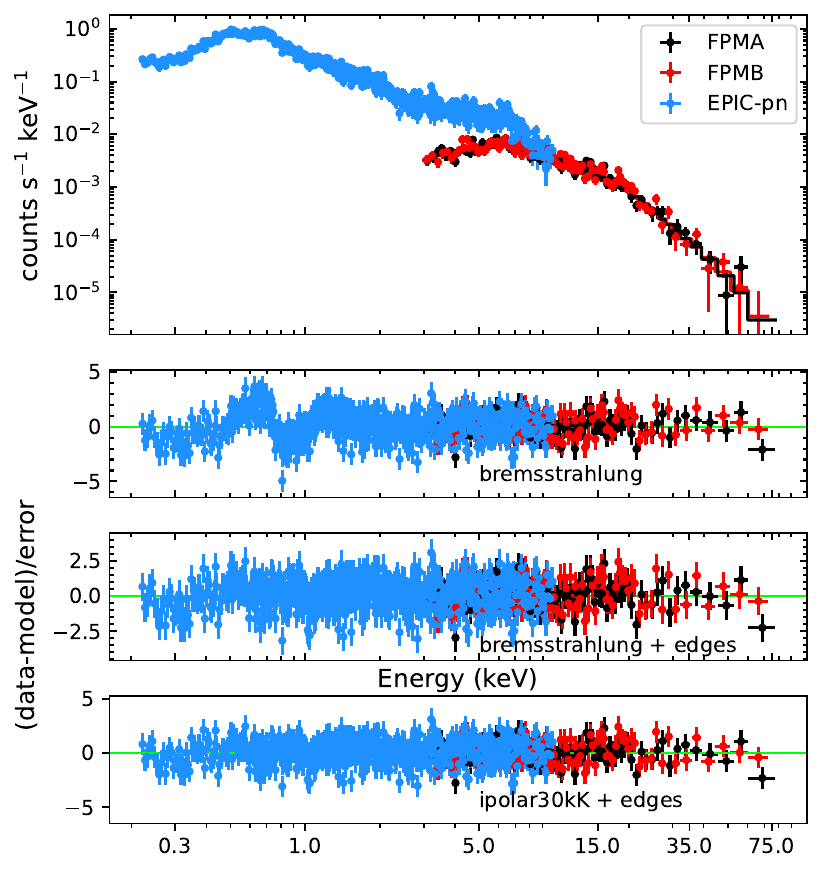}
\caption{XMM-Newton/PN and NuSTAR spectra of V1674 Her obtained on 2023 October 14 (top panel). Middle panels show residuals of the \texttt{bremsstrahlung} model fit without and including the O VII and O VIII absorption edges. Bottom panel shows residual plot for \texttt{ipolar30kK} model including the aforementioned absorption edges.}
   \label{fig:bremss}\label{fig:NuXMM_spec}
   \end{figure*}

We first fit the \xmm EPIC-pn (0.2-10\,keV) and \nustar (3-78\,keV) NuSTAR spectra with a composite model: 
\texttt{tbabs*pcfabs*(reflect*bremss + bbody+gauss)}.
The \texttt{tbabs} model \citep{2000ApJ...542..914W} accounts for Galactic absorption, while \texttt{pcfabs} stands for a partially covered absorber describing complex intrinsic absorption in the IP. Some of the radiation from the shock accretion column is expected to reflect off the white dwarf surface. 
This component was first detected unambiguously in hard X-ray spectra of the intermediate polars V709~Cas, NY~Lup, and V1223~Sgr \citep{2015ApJ...807L..30M}. We used the Xspec convolution model \texttt{reflect} \citep{1995MNRAS.273..837M} to describe the contribution of reflected photons to the continuum spectrum. We simply adopt the $67^\circ$ binary inclination angle for \nova \citep{2024MNRAS.527.1405H} as phase averaged observing angle and subsequently fix the value of \texttt{cosIncl} parameter in the \texttt{reflect} model to 0.39. A gaussian emission component at 6.4\,keV was also included to represent emission from the neutral Fe$-k_{\alpha}$ line.
A residual feature around 0.7-0.9\,keV  (Figure~\ref{fig:bremss}, panel 2) can be removed by adding absorption edges from H-like and He-like oxygen ions (O VII and O VIII) at 0.74\,keV and 0.87\,keV (Figure~\ref{fig:bremss}, panel 3).
Some fraction of these ionized absorber features could originate from the ejecta of the recent nova outburst. The significance of each of these model components (F-test) is evaluated and summarized in Table~\ref{table:2}, and the best-fit parameters are reported in Table~\ref{table:3}.

\section{White dwarf Mass and Magnetic Field Estimation}
\label{sec:wme}
As shown in Section~\ref{spc}, the hard X-ray emission from intermediate polars can be described by bremsstrahlung emission from shocked gas falling onto the surface of the white dwarf. The temperature of the post-shock region is related to the compactness of the white dwarf \citep{1977ApJ...215..265K}. However, inferring the shocked plasma temperature with a thermal bremsstrahlung fit underestimates the peak shock temperature because the post-shock region has a temperature gradient and is not isothermal. The post-shock region emission model \citep[\texttt{ipolar30kk}, ][]{2025A&A...700A.180S} calculates hard X-ray spectra as a function of white dwarf mass ($M$) and magnetosphere radius ($R_\text{M}$) in a two-parameter grid, considering a realistic accretion flow profile for the falling gas and the finite height of the magnetosphere. 
The \texttt{ipolar30kk} model implements an  $M\!-\!R$ relation based on direct interpolation of numerical $M\!-\!R$ relations from white dwarf models from \citet{2001PASP..113..409F} with a thick hydrogen envelope and a temperature of 30,000\,K (see \citealt{2025A&A...700A.180S} for a more detailed description of the \texttt{ipolar30kk} model). The resulting spectral models have some degeneracy with respect to the two free model parameters $M$ and $R_\text{M}$.  Thus, an additional constraint on magnetosphere radius is required to remove the model degeneracy and infer the mass of the white dwarf from X-ray spectral fits. For accretion to occur, the magnetosphere radius has to be smaller or equal to the co-rotation radius $R_\text{C}$ at which the Keplerian angular velocity of the accreting material equals the angular velocity of the white dwarf spin \citep{2002apa..book.....F}.
Under the assumption of an accretion disk truncated at the co-rotation radius  
\begin{gather}
\label{eq:2}
    R_\text{M}\approx R_\mathrm{C}= {(GMP^2_\text{spin}\mathbin{/}4\pi^2)}^{1/3}
\end{gather}
we modeled the contemporaneous \xmm and \nustar spectra of \nova in quiescent state with \texttt{tbabs*pcfabs*edge*edge*(reflect*ipolar30kk+
bbody+gauss)} model. We fixed the gaussian line energy and width at 6.4\,keV and 0.10\,keV to model the neutral Fe-K$_\alpha$ line emission. The details and results of this model fit are recorded in Table~\ref{table:3}, and yield a best-fit value of $M=1.09^{+0.07}_{-0.06}$\,$M_{\odot}$ for the white dwarf mass. The residuals in Figure \ref{fig:bremss} and $\chi^2$/(d.o.f) value of 816.4/781 indicate that this model is a good description of the observed spectra.

\begin{table*}[t]
\centering
\begin{threeparttable}
\caption{Best fit spectral parameters for contemporaneous XMM-Newton and NuSTAR spectrum of V1674 Her obtained on 14 October 2023. The spectral models are plotted in Figure \ref{fig:bremss} and described in sections \ref{sec:spcbremss}, \ref{sec:wme} and \ref{sec:wme2}.} 
\label{table:3}
\begin{tabular}{ l l c c c }
\hline
Model Component & Parameters & \texttt{bremss} & \texttt{ipolar30kK} & \texttt{ipolar30kK} \\
 &  &  & spectral$^d$ & spectral+timing$^e$ \\
\hline
\texttt{tbabs} & $n\text{H}$ [$10^{22}\,\text{cm}^{-2}$]$^a$ & $0.30^{+0.02}_{\dagger}$ & $0.32^{+0.02}_{-0.02}$ & $0.32^{+0.02}_{-0.02}$ \\
$\texttt{edge}_{1}$ & $E_{1}^*$ & 0.74 & 0.74 & 0.74 \\
& $\tau_{1}$ & $0.92^{+0.06}_{-0.06}$ & $0.92^{+0.06}_{-0.06}$ & $0.92^{+0.06}_{-0.06}$ \\
$\texttt{edge}_{2}$ & $E_{2}^*$ & 0.87 & 0.87 & 0.87 \\
& $\tau_{2}$ & $0.55^{+0.06}_{-0.06}$ & $0.52^{+0.06}_{-0.06}$ & $0.53^{+0.06}_{-0.06}$ \\
\texttt{pcfabs} & $n\text{H}$ [$10^{22}\,\text{cm}^{-2}$]$^b$ & $73.75^{+13.54}_{-10.34}$ & $59.06^{+8.80}_{-7.14}$ & $57.48^{+8.90}_{-6.88}$ \\
& $f$ & $0.48^{+ 0.06}_{-0.07}$ & $0.52^{+ 0.06}_{-0.03}$ & $0.52^{+ 0.06}_{-0.06}$ \\
\texttt{reflect} & Scale & 1.0 & 1.0 & 1.0 \\
& Abund. & $1.02^{+0.92}_{-0.44}$ & $0.68^{+0.48}_{-0.27}$ & $0.74^{+0.48}_{-0.28}$ \\
& Abund.-Fe$^c$ & 1.02 & 0.68 & 0.74 \\
& $\cos{i}^*$ & 0.39 & 0.39 & 0.39 \\
\texttt{polar} & fall-height [$R$] & -- & 20.47 & $11.45^{+1.58}_{-1.30}$ \\
& Mass [$M_\odot$] & -- & $1.09^{+0.07}_{-0.06}$ & $1.12^{+0.06}_{-0.06}$ \\
& $K$ [$10^{-30}$] & -- & $9.62^{+5.33}_{-3.70}$ & $8.63^{+5.03}_{-3.35}$ \\
\texttt{bremss} & $kT$ [keV] & $27.35^{+4.09}_{-3.19}$ & -- & -- \\
& $K$ [$10^{-4}$] & $7.54^{+1.04}_{-0.87}$ & -- & -- \\
\texttt{bbody} & $kT$ [keV] & $0.124^{+0.003}_{-0.003}$ & $0.120^{+0.003}_{-0.003}$ & $0.121^{+0.003}_{-0.003}$ \\
& $K$ [$10^{-4}$] & $1.89^{+0.44}_{-0.35}$ & $2.33^{+0.56}_{-0.44}$ & $2.31^{+0.56}_{-0.44}$ \\
$\texttt{gauss}$ & $E_{1}^*$ & 6.4 & 6.4 & 6.4 \\
& $\sigma$ & $0.10^{+0.06}_{-0.06}$ & $0.10^*$ & $0.10^*$ \\
& $K$ [$10^{-6}$] & $5.31^{+1.58}_{-1.36}$ & $4.58^{+1.19}_{-1.15}$ & $4.56^{+1.19}_{-1.15}$ \\
\texttt{constant} & \nustar FPMA & $1.17^{+0.04}_{-0.04}$ & $1.18^{+0.04}_{-0.04}$ & $1.18^{+0.04}_{-0.04}$ \\
& \nustar FPMB & $1.24^{+0.04}_{-0.04}$ & $1.25^{+0.04}_{-0.04}$ & $1.25^{+0.04}_{-0.04}$ \\
& \xmm $pn^*$ & 1.0 & 1.0 & 1.0 \\
\hline
$\chi^2/\text{d.o.f.}$ & & 821.1/780 & 816.4/781 & 891.20/847 \\
\hline
\end{tabular}
\begin{tablenotes}\small
\item $K$: normalization; $f$: dimensionless covering fraction $(0<f\leq 1)$; $\tau$: absorption depth; $^a$ Galactic absorption; $^b$ intrinsic absorption; $^c$ Fe abundance tied to the overall abundance; $^d$ $R_\text{M}=R_\text{C}$; $^e$ joint fit with the timing power spectrum.
\item $^*$ Fixed, $^{\dagger}$ Parameter pegged at lower/upper limit
\end{tablenotes}
\end{threeparttable}
\end{table*}

\subsection{Mass constraints from a frequency-domain analysis of the X-ray light curve}\label{sec:wme2}
\begin{figure*}[tb]
  \centering
  \includegraphics[scale=0.5]{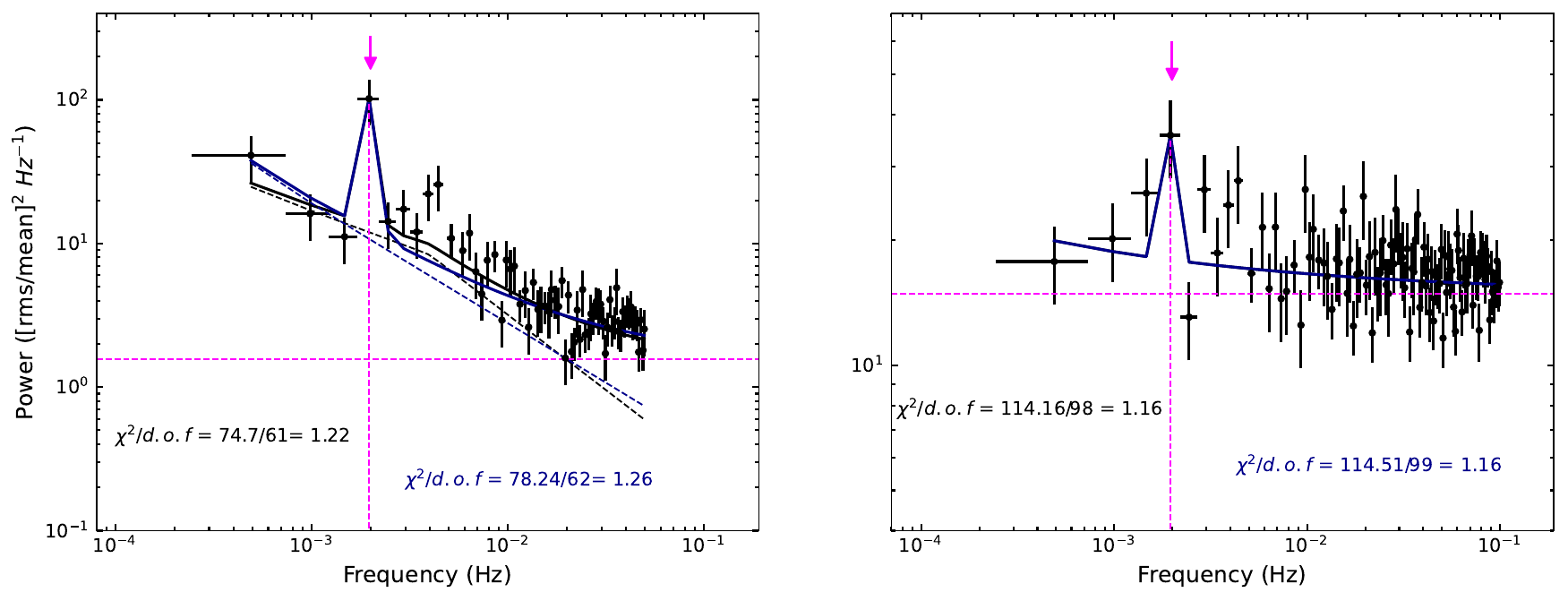}
\caption{Power spectra of XMM-Newton/PN (left) and NuSTAR (right) light curves of V1674 fitted with Gaussian + Constant + Single power-law (blue) and   Gaussian + Constant + Broken power-law (black). Vertical magenta arrow (WD spin frequency), vertical magenta line (Gaussian component at WD spin frequency) and horizontal magenta line (white noise level). }
   \label{fig:pow_spec}
 \end{figure*}

With an estimated mass of $M\sim1.09\,M_{\odot}$, \nova is more massive than the average white dwarf. The mass distribution of isolated white dwarfs peaks at $\sim0.6\,M_{\odot}$ \citep{1990ARA&A..28..103W}. White dwarfs in cataclysmic variables, however, are more massive on average \citep[$\langle M \rangle\sim 0.83\,M_{\odot}$,][]{2020MNRAS.494.3799P}. Magnetic isolated white dwarfs are also more massive than their non-magnetic counterparts \citep[e.g.,][]{2022ApJ...935L..12B}. 
Fixing the scale height of the magnetosphere at the corotation radius, as assumed in Section~\ref{sec:wme}, can lead to an underestimation of the white dwarf mass. 
An independent constrain on the white dwarf mass can be obtained using power spectral (Fourier-domain) analysis of the X-ray light curve, interpreting the break frequency as the dynamical timescale at the magnetospheric radius.
The aperiodic X-ray and optical frequency power spectra of several intermediate polars show a break in the power-law spectrum at a frequency $\nu_{b}$ assumed to correspond to the frequency of a Keplerian orbit at the magnetosphere radius \citep[see, e.g., ][]{2010A&A...513A..63R,2019MNRAS.482.3622S} 
\begin{gather}
\label{eq:4}
    \nu_{b}=\frac{1}{2\pi}\sqrt{\frac{GM}{R_M^3}}
\end{gather}
We extracted separate frequency power spectra for 10\,s binned {\it XMM}/EPIC-pn (0.3-10\,keV) and the combined \nustar FPMA and FPMB (3-78\,keV) light curves presented in Figure~\ref{fig:lc_xmm_Nu}.
We modeled the resulting power spectra with a (single or broken) power law, a Gaussian component at the white dwarf spin frequency of 1/501.72\,s \citep{2021ApJ...922L..42D}, and a constant white noise level, as shown in Figure~\ref{fig:pow_spec}. The \nustar power spectrum is primarily white noise and cannot be used to derive a break frequency. 
The power spectrum of the \xmm light curve shows evidence for a break frequency at $\nu_b=3.95$\,mHz (Figure~\ref{fig:pow_spec}).
We tested the significance of the spectral break by a broken power law model ($\chi^2/\mathrm{d.o.f.} = 74.7/61$) with a single power law ($\chi^2/\mathrm{d.o.f.} = 78.2/62$).
The resulting $F$-statistic of 2.89 corresponds to a $p$-value of 0.09, indicating that the improvement provided by the broken power-law model is not statistically significant at the $>95\%$ confidence level.

Even if the observed spectral break is not significant, we proceeded to estimate the white dwarf mass by doing a joint fit of the \xmm EPIC-pn Fourier power spectrum (converted to \texttt{Xspec} format) and the energy spectra collected by \xmm and \nustar. 
The joint fit yields a white dwarf mass estimate of $M=1.12\pm{0.06}$\,$M_{\odot}$. This value is compatible to the $1.09$\,$M_{\odot}$ estimate obtained in the previous section under the assumption of co-rotation radius for magnetosphere size. 

\begin{figure*}[tb]
  \centering
  \includegraphics[scale=0.7]{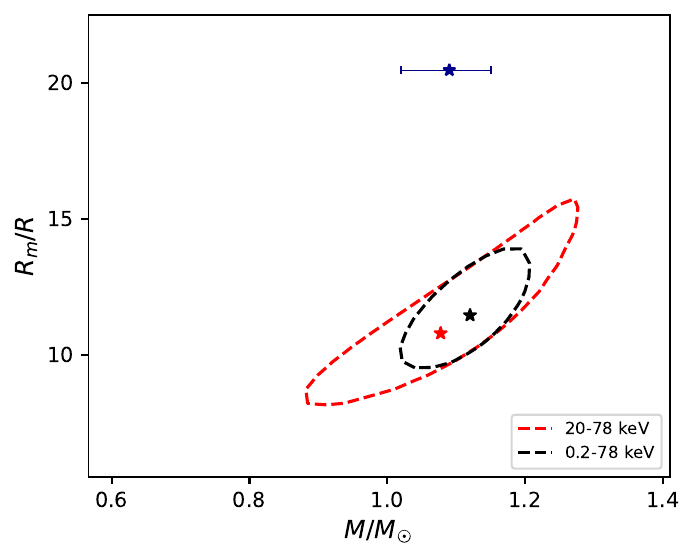}
\caption{Contour plots of the 1$\sigma$ confidence regions from joint fits of the power spectra and 20-78 keV (red) and 0.2-78 keV (black) energy spectrum. The blue star symbol and error bars indicate the white dwarf mass and 1$\sigma$ confidence interval obtained under the assumption of co-rotation radius for magnetosphere.}
   \label{fig:conf_reg}
\end{figure*}

To minimize systematic uncertainties associated with an incomplete description of the complex absorption, emission, and reflection components that are present at energies $\leq 20$\,keV, we repeated the joint spectral fit, but using only the NuSTAR 20-78\,keV data band for the energy spectrum and modeled the energy spectrum part as an unabsorbed post-shock region emission (\texttt{ipolar30kk+bknpow+gauss+powerlaw}).
Discarding the low-energy data yields less constraining results, with $M=1.08^{+0.14}_{-0.14}$ $M_{\odot}$ ($\chi^2/d.o.f. = 95.7/93$), but show that the derived white dwarf mass values are robust to changes in the model describing the features of the lower-energy X-ray spectra (see also Figure \ref{fig:conf_reg}).

\subsection{Estimate of the white dwarf magnetic field}\label{sec:B}
Using  contemporaneous \xmm and \nustar observations of \nova, the 0.2-78\,keV unabsorbed flux is estimated as 
$F_X = (1.82 \pm {0.01})\times10^{-11}$\,\fluxcgs. Assuming a source distance of $D=6.3^{+3.8}_{-2.4}$\,kpc from \cite{2023MNRAS.521.5453S}, this translates to an X-ray luminosity of $L_X=4\pi D^2F_X = 8.7^{+10.4}_{-6.6}\times10^{34}$\,\lumcgs,  indicating that \nova is indeed a luminous intermediate polar. 
In intermediate polars, the X-ray luminosity can be estimated as \citep{2019MNRAS.482.3622S}
\begin{gather}
\label{eq:7}
    L_X=GM\dot{M}\left(\frac{1}{R}-\frac{1}{R_\text{M}}\right)
\end{gather}
from which we can derive a mass accretion rate of $\dot{M}=3.0^{+3.6}_{-2.3}\times10^{17}$\,g\,s$^{-1}$. 
The white dwarf magnetic field $B$ can then be estimated by equating the ram pressure ($\rho v^2$) and magnetic pressure ($B^2/8\pi$) at the Alfv\'{e}n radius $R_{\rm M}$.
In the case of disc accretion, the magnetic field becomes
\begin{gather}
\label{eq:8}
    B^4=\frac{2\,GM\dot{M}^2}{R^5}\left(\frac{R_\text{M}}{\xi R}\right)^7
\end{gather}
with $\xi=0.5$ \citep{2019MNRAS.482.3622S, 2002apa..book.....F}. 
Substituting in the relevant values we obtain $B = 21.3^{+6.6}_{-5.7}\,(\mathrm{stat})^{+12.9}_{-8.1}\,(\mathrm{sys})\,\mathrm{MG}$, where the systematic uncertainty is associated to the uncertainty in the distance to \nova.

We note that this value lies at or above the upper end of the typical $0.1$--$10$\,MG magnetic field range for intermediate polars \citep{2008MNRAS.387.1157R}, suggesting that V1674 Her may be an unusually magnetic system, potentially approaching the regime of polars. Alternatively, if cyclotron cooling contributes significantly in the post-shock region, the inferred shock temperature, and hence the white dwarf mass, may be underestimated, which would in turn increase the derived magnetic field strength to even higher values.

\section{Implications for white dwarf mass estimates in fast novae} \label{sec:xlc}
The optical decline times $t_2$ and $t_3$ (defined as the times required for the
nova to fade by 2 and 3 magnitudes from the optical peak, see Figure~\ref{fig:lc1}) provide a simple proxy for the mass-loss timescale and, by extension, the white dwarf mass \citep{1981ApJ...243..926S}. 
\citet{2025ApJ...993..232S} calculates an empirical relation between $\log t_3$ and $M$ by fitting a linear regression to the mass estimates and decline times of $\sim$190 novae, obtaining
\begin{gather}
\label{eq:7}
    t_3=1,900\,\text{d} \times 10^{-1.73\,M}
\end{gather}
For \nova, we estimate $t_3=(2.45 \pm 0.98)\,\text{d}$ from a polynominal fit to the optical light curve (Figure~\ref{fig:lc1}). Inverting Equation~\ref{eq:7}, we derive a mass estimate of $1.67\pm0.10\,M_{\odot}$. This value exceeds not only our estimate of the mass from X-ray spectroscopy but also the Chandrasekhar limit, indicating that factors other than the white dwarf mass are likely to play a significant role in shaping the evolution of the optical light curve in extreme cases like the fast nova \nova. Similar, though less pronounced, deviations from the linear regression are also seen for a few outliers in \citet{2025ApJ...993..232S}.

\citet{2021ApJ...922L..42D} estimated a lower limit of
$M > 1.05\,M_{\odot}$ for \nova\ based on the detection of strong
optical neon emission lines \citep{2021ATel14746....1W}, which suggest that the
underlying white dwarf is of the oxygen-neon (ONe) type (see also \citealt{2015MNRAS.446.2599D}).
Since ONe white dwarfs are expected to originate from more massive progenitors,
they are typically associated with higher masses ($\gtrsim 1.05$--$1.1\,M_\odot$).
Our direct estimates indicate $M \gtrsim 1.09^{+0.07}_{-0.06}\,M_{\odot}$, placing \nova in the ultra-massive category \citep[see][for definitions and classifications]{2021A&A...646A..30A}.
To put this measurement in perspective, 
white dwarf masses within 100\,pc in the SDSS sample show a bimodal distribution with peaks at $\sim0.6$ and $\sim0.8\,M_{\odot}$ \citep{2018MNRAS.479L.113K},
whereas intermediate polars in the {\it Swift}-BAT sample peak at $\sim0.82\,M_{\odot}$ \citep{2025A&A...700A.180S}. 
With $\sim1.09$ $M_{\odot}$, \nova is placed at the high end tail part of both the SDSS and the {\it Swift}-BAT sample distributions. Although both samples are important for our mass comparison, the {\it Swift}-BAT sample is more relevant as it contains only white dwarfs in intermediate polar systems and the mass is calculated using the same X-ray spectroscopy method.

\subsection{Is \nova a potential Type Ia supernova candidate?}
With an estimated mass of $1.09$--$1.12\,M_{\odot}$, \nova is well below the
Chandrasekhar limit ($\sim1.44\,M_{\odot}$) and is therefore not an imminent
Type Ia supernova progenitor. Even under the extreme assumption of continuous accretion with no mass loss, its accretion rate of $\dot{M}\sim4.9\times10^{-9}\,M_{\odot}\,\text{yr}^{-1}$ implies a growth timescale of $\gtrsim 6\times10^{7}$\,years to reach $M_{\rm Ch}$.

\begin{figure}[h]
  \centering
  \includegraphics[scale=0.6]{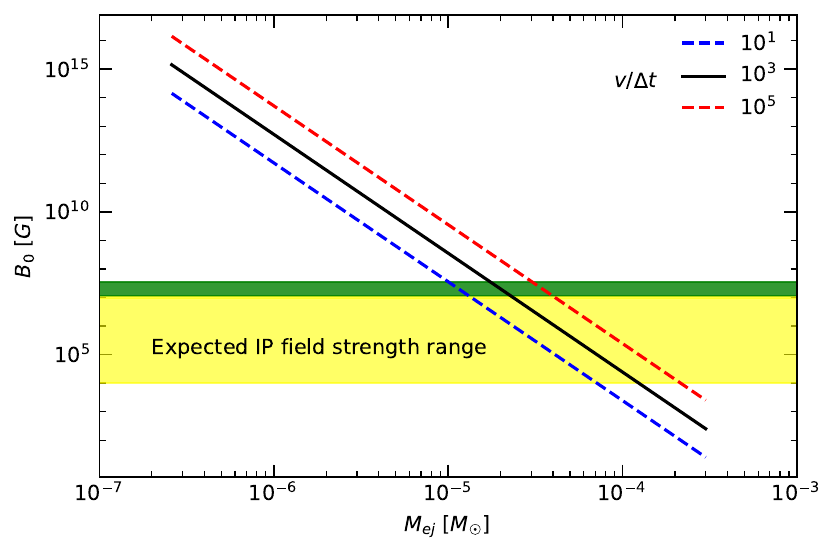}
\caption{White dwarf magnetic field B as a function of ejecta mass and $v/\Delta t$, adapted from  \cite{2021ApJ...922L..42D}. Horizontal green band indicates our estimate for V1674 Her's magnetic field strength.}
   \label{fig:B}
\end{figure}

In reality, nova eruptions expel a significant fraction of the accreted material. 
In \nova, the observed fractional spin-down of the white dwarf of 
$\Delta P/P \sim 2\times10^{-4}$ \citep{2021ApJ...922L..42D} before and after the 2021 eruption can be interpreted as the loss of angular momentum
 carried by magnetically coupled ejecta. Assuming a dipolar field with
$B = 21.3^{+6.6}_{-5.7}\,(\mathrm{stat})^{+12.9}_{-8.1}\,(\mathrm{sys})\,\mathrm{MG}$ (Section~\ref{sec:B}), this implies an ejecta mass of $M_\text{ej}\sim 9\times10^{-6}-3\times10^{-5}\,M_{\odot}$(Figure~\ref{fig:B}). To estimate the fraction of accreted material retained by the white dwarf, we first determine the accreted mass $M_\text{acc}$ required to ignite the thermonuclear runaway. This can be done by considering the critical pressure $P_\text{crit}$ needed for ignition: 
\begin{equation}
   \label{eq:9}
    P_\text{crit}=\frac{GMM_\text{acc}}{4\pi R^4}
\end{equation}
where $P_\text{crit} \sim 10^{19}-10^{20}\,\text{dyn}\,\text{cm}^{-2}$ \citep{1982ApJ...257..752F,2025A&A...698A.251J}. This implies $M_\text{acc} \sim 2\times10^{-4}\,M_{\odot}$ for \nova, indicating that $4-15\%$ of the total accreted mass is retained after each nova outburst cycle, resulting on a gradual but slow growth of the white dwarf. Under this scenario, \nova will not reach the Chandrasekhar mass threshold for another $> 10^8 - 10^9$\,years. 

\section{Summary and conclusions} \label{sec:cnc}

The extremely rapid optical decline, ultra-fast expansion velocities, and classification of \nova\ as an oxygen-neon nova all point to the presence of a high-mass white dwarf. In particular, the exceptionally short decline time ($t_2 \sim 1$ day) would, under standard interpretations, suggest a white dwarf mass close to the Chandrasekhar limit \citep[e.g.,][]{2005ApJ...623..398Y,2016ApJ...819..168H}. In this work, however, we directly constrain the white dwarf mass using X-ray spectroscopy by exploiting the relation between the post-shock temperature and the compactness of intermediate polars \citep[e.g.,][]{1977ApJ...215..265K}. 

By modeling contemporaneous \xmm\ EPIC-pn and \nustar\ spectra obtained during quiescence with a physically motivated post-shock accretion column model, we derive a mass of $M \sim 1.1\,M_{\odot}$. This result confirms that \nova\ hosts a relatively massive white dwarf, but one that is significantly below the Chandrasekhar limit. The discrepancy between this direct X-ray constraint and the higher masses inferred from empirical decline-time relations suggests that such relations may overestimate white dwarf masses in extremely fast novae, and that additional parameters (accretion rate, magnetic field strength, or envelope properties) may play a key role in determining nova timescales \citep{2005ApJ...623..398Y,2016ApJ...819..168H}. 

We also estimate a surface magnetic field strength of $B \sim 21$\,MG for \nova. This value lies at or above the upper end of the typical $0.1$--$10$\,MG range for intermediate polars \citep{2008MNRAS.387.1157R,2017PASP..129f2001M}, suggesting that \nova\ may be an highly magnetized system, potentially approaching the regime of polars \citep[e.g.,][]{2004ApJ...614..349N,2015SSRv..191..111F}. 

Cyclotron cooling is generally negligible in intermediate polars \citep{2016A&A...591A..35S}. However, the relatively high magnetic field inferred for \nova\ introduces a systematic uncertainty in our results. If cyclotron cooling contributes significantly to the post-shock region and is not included in the modeling of the X-ray spectra, the inferred mass of the white dwarf may be underestimated \citep[e.g.,][]{1994ApJ...426..664W}, which implies that the true mass could be somewhat higher and approach the Chandrasekhar limit. 

The \texttt{ipolar30kk} model includes thermal bremsstrahlung emission but does not account for cyclotron cooling. In contrast, models such as \texttt{mcvspec} \citep{2023ApJ...954..138V,2025ApJ...987...53F} incorporate cyclotron emission and reflection from the stellar surface self-consistently. Adopting \texttt{mcvspec} would increase the inferred white dwarf mass due to the inclusion of cyclotron cooling, which is negligible in intermediate polars but important in polar systems.
However, \texttt{mcvspec} adopts the $M$–$R$ relation from \citet{1972ApJ...175..417N} rather than the more realistic prescription used in \texttt{ipolar30kk}, which includes finite-temperature effects \citep{2025A&A...700A.180S}. In addition, \texttt{mcvspec} does not explicitly incorporate the gravitational potential or the dipolar magnetic field geometry of the post-shock region. These simplifications can introduce additional systematic uncertainties in the inferred mass estimates \citep{1999MNRAS.306..684C,2007MNRAS.379..779S}.

The magnetic field inferred from X-ray spectroscopy of \nova is relatively strong for an intermediate polar. Infrared observations could provide independent constraints. In particular, cyclotron humps in the infrared spectrum could provide a robust and model-independent estimate of magnetic field strength in polars and potentially in systems like \nova \citep[e.g.,][]{1993MNRAS.262..285F}. 
Such observations will be essential to confirm the magnetic nature of this system and to further constrain the role that magnetic fields play in shaping the accretion flow and outburst properties of fast novae.

\section{Acknowledgments} \label{sec:xlc}
The authors thank the anonymous referee useful feedback that improved this manuscript. We thank professors Mike Nowak and Jim Buckley for insightful discussions that contributed to the early development of this work. We also thank professor Robert Quimby for providing the Evryscope and MLO-ASC optical data. This work was based on observations obtained with \xmm, an ESA science mission with instruments and contributions directly funded by ESA Member States and NASA. This research has used data from the \nustar mission, a project led by the California Institute of Technology, managed by the Jet Propulsion Laboratory and funded by the National Aeronautics and Space Administration. Data analysis was performed using the \nustar Data Analysis Software (\texttt{NuSTARDAS}), jointly developed by the ASI Science Data Center (SSDC, Italy) and the California Institute of Technology (USA). We acknowledge the use of public data from
the \swift data archive. We acknowledge with thanks the variable star observations from the AAVSO International Database contributed by observers worldwide and used in this research. 
\clearpage
\bibliography{v1674Her}{}
\bibliographystyle{aasjournal}
\end{document}